\documentclass[11pt,fleqn]{article}
\usepackage{graphicx}

\begin{document}

\title{
{\sf Pad\'e-Improved Extraction of  $\alpha_s(M_\tau)$ from $R_\tau$} 
}

\author{
T.G. STEELE \\
{\sl Department of Physics and Engineering Physics and}\\
{\sl Saskatchewan Accelerator Laboratory}\\
{\sl University of Saskatchewan}\\
{\sl Saskatoon, Saskatchewan S7N 5C6, Canada.}
\and
{V. ELIAS}\\
{\sl Department of Applied Mathematics,} 
{\sl The University of Western Ontario,} \\
{\sl London, Ontario N6A 5B7, Canada. }
}
\maketitle
\begin{abstract}
The perturbative series
used to extract $\alpha_s\left(M_\tau\right)$
from the $\tau$ hadronic width 
exhibits slow convergence.  
Asymptotic Pad\'e-approximant and Pad\'e summation techniques  provide an estimate 
of these unknown higher-order effects, leading to 
values for $\alpha_s\left(M_\tau\right)$ that are about
$10\%$ smaller than 
current estimates.  Such a reduction improves the agreement of 
$\alpha_s\left(M_\tau\right)$  with the 
QCD evolution of the strong coupling constant from $\alpha_s\left(M_z\right)$.
\end{abstract}
The Particle Data Group (PDG) quotes the following values for the
strong coupling constant as determined from  $Z^0$ and $\tau$ decays \cite{pdg98}.
\begin{eqnarray}
& &\alpha_s\left(M_\tau\right)=0.35\pm 0.03
\label{alpha_tau_pdg}
\\
& &\alpha_s\left(M_Z\right)=0.119\pm 0.002
\label{alpha_Z_pdg}
\end{eqnarray}
Since these determinations of $\alpha_s$ occur at such widely separated energies, 
the compatibility of these values of $\alpha_s$ with the QCD evolution of the coupling constant
is an important test of both QCD and the phenomenological results used to extract the coupling constant
from the experimental data.  In particular, $\alpha_s\left(M_\tau\right)$ is sufficiently large
that presently unknown terms from higher order perturbation theory could 
substantially alter the
value of $\alpha_s\left(M_\tau\right)$  extracted from the experimental data.
Pad\'e approximant methods provide estimates of the aggregate effect of
(presently unknown) terms from higher-order perturbation theory \cite{pade,samuel2,apap,samuel}.
As shown below, the use of Pad\'e summation
to estimate such terms 
leads to a  decrease in the
value of $\alpha_s\left(M_\tau\right)$ extracted from $\tau$ decays, improving the 
compatibility of $\alpha_s\left(M_\tau\right)$ and $\alpha_s\left(M_Z\right)$ with the QCD evolution of the
coupling constant.

The QCD evolution of the coupling constant is governed by the $\beta$ function
which is now known to 4-loop order \cite{beta}.  Using the conventions of
\cite{chetyrkin1}, $a\equiv\frac{\alpha_s}{\pi}$ satisfies the differential equation
\begin{eqnarray}
& &\mu^2\frac{d a}{d \mu^2}=
\beta(a)=
-a^2\sum_{i=0}^\infty \beta_i a^i\quad ,\quad a\equiv\frac{\alpha_s}{\pi}
\label{evolution}
\\
& &\beta_0=\frac{11-\frac{2}{3}n_f}{4}
\quad ,\quad
\beta_1=\frac{102-\frac{38}{3}n_f}{16}\quad ,\quad
\beta_2=\frac{\frac{2857}{2}-\frac{5033}{18}n_f+\frac{325}{54}n_f^2}{64}
\\[5pt]
& &
\beta_3=114.23033-27.133944 n_f+1.5823791 n_f^2-5.8566958\times 10^{-3}n_f^3
\end{eqnarray}
Using the value  $\alpha_s\left(M_z\right)$ as an initial condition, the coupling
constant can be evolved to the desired energy using the differential equation
(\ref{evolution}).  The only subtlety in this approach is the location of flavour thresholds
where the number of effective flavour degrees of freedom $n_f$ change.  In general, matching conditions
must be imposed at these thresholds to relate QCD with   $n_f$ quarks to an effective theory
with $n_f-1$ light quarks and a decoupled heavy quark \cite{rodrigo}.  Using the matching threshold $\mu_{th}$ defined by
$m_q(\mu_{th})=\mu_{th}$, where $m_q$ is the running quark mass, the matching condition to three-loop order is 
\cite{chetyrkin2}
\begin{equation}
a^{(n_f-1)}\left(\mu_{th}\right)=
a^{(n_f)}\left(\mu_{th}\right)\left[1+0.1528 \left[a^{(n_f)}\left(\mu_{th}\right)\right]^2
+\left\{0.9721-0.0847\left(n_f-1\right)\right\}\left[a^{(n_f)}\left(\mu_{th}\right)\right]^3\right]
\label{matching}
\end{equation}
leading to a discontinuity of $\alpha_s$ across the threshold.
Thus to determine the coupling constant at energies between the $c$ quark threshold and the $b$ quark threshold,
the $\beta$ function with $n_f=5$ is used to run $\alpha_s^{(5)}$ from $M_Z$ to $\mu_{th}=m_b(\mu_{th})\equiv m_b$ using
(\ref{alpha_Z_pdg}) as an initial condition.  The matching condition
(\ref{matching}) is then imposed to find the value of $\alpha_s^{(4)}(m_b)$ which is then used as an initial condition
to evolve $\alpha_s$ to lower energies via the $n_f=4$  beta   function.

If $\alpha_s\left(M_Z\right)$ is used as the input value to determine
the QCD prediction of $\alpha_s\left(M_\tau\right)$, then  one might legitimately be concerned about the effect of
(unknown) higher-order terms in the $\beta$ function at lower energies where $\alpha_s$ is larger.
Pad\'e approximations have proven their utility in determining higher-order terms in the $\beta$ function.
For example, using as input the four-loop $\beta$ function in $N$-component massive $\phi^4$
scalar field theory \cite{O(N)},
asymptotic Pad\' e methods
described in Section II of \cite{apap}
are able to
predict the five-loop term to better than $10\%$ of the known five-loop contributions for $N\le 4$ 
\cite{samuel2,elias,chishtie}.  
When these same methods are applied to QCD, the following predictions for the unknown five-loop contribution to the $\beta$
function are obtained \cite{elias}.
\begin{eqnarray}
n_f=4: & & \beta_4=83.7563
\label{beta_5_nf=4}
\\
n_f=5: & & \beta_4=134.56
\label{beta_5_nf=5}
\end{eqnarray}
From these predictions, $\beta$ functions containing $[2\vert 2]$ Pad\'e approximants
can be constructed to estimate the sum of all higher-order contributions.
These Pad\'e-summations, whose Maclaurin expansions reproduce $\beta_1,~\beta_2,~\beta_3$ and
and the asymptotic Pad\'e-approximant estimates (\ref{beta_5_nf=4},\ref{beta_5_nf=5}) of
$\beta_4$, are given by:
\begin{eqnarray}
n_f=4: & & \beta(a)=-\frac{25x^2}{12}\left[\frac{1-5.8963 a-4.0110 a^2}{1-7.4363 a+4.3932 a^2}\right]
\label{beta_pade_nf4}
\\[5pt]
n_f=5: & & \beta(a)=-\frac{23x^2}{12}\left[\frac{1-5.9761 a-6.9861 a^2}{1-7.2369 a-0.66390 a^2}\right]
\label{beta_pade_nf5}
\end{eqnarray}
Thus the QCD prediction of $\alpha_s\left(M_\tau\right)$ depends on only two parameters: the initial condition
$\alpha_s\left(M_z\right)$ and the position of the five-flavour threshold defined by $m_b(m_b)=m_b$.  
As will be discussed below, the uncertainty in the Particle Data Group value \cite{pdg98} for this threshold
\begin{equation}
4.1\, {\rm GeV}\le m_b(m_b)\le 4.4\,{\rm GeV}
\label{mb_pdg}
\end{equation}
has a negligible effect on the QCD prediction of $\alpha_s\left(M_\tau\right)$ compared with the
uncertainty in $\alpha_s\left(M_z\right)$ (\ref{alpha_Z_pdg}).

\begin{figure}
\centering
\includegraphics[scale=0.7]{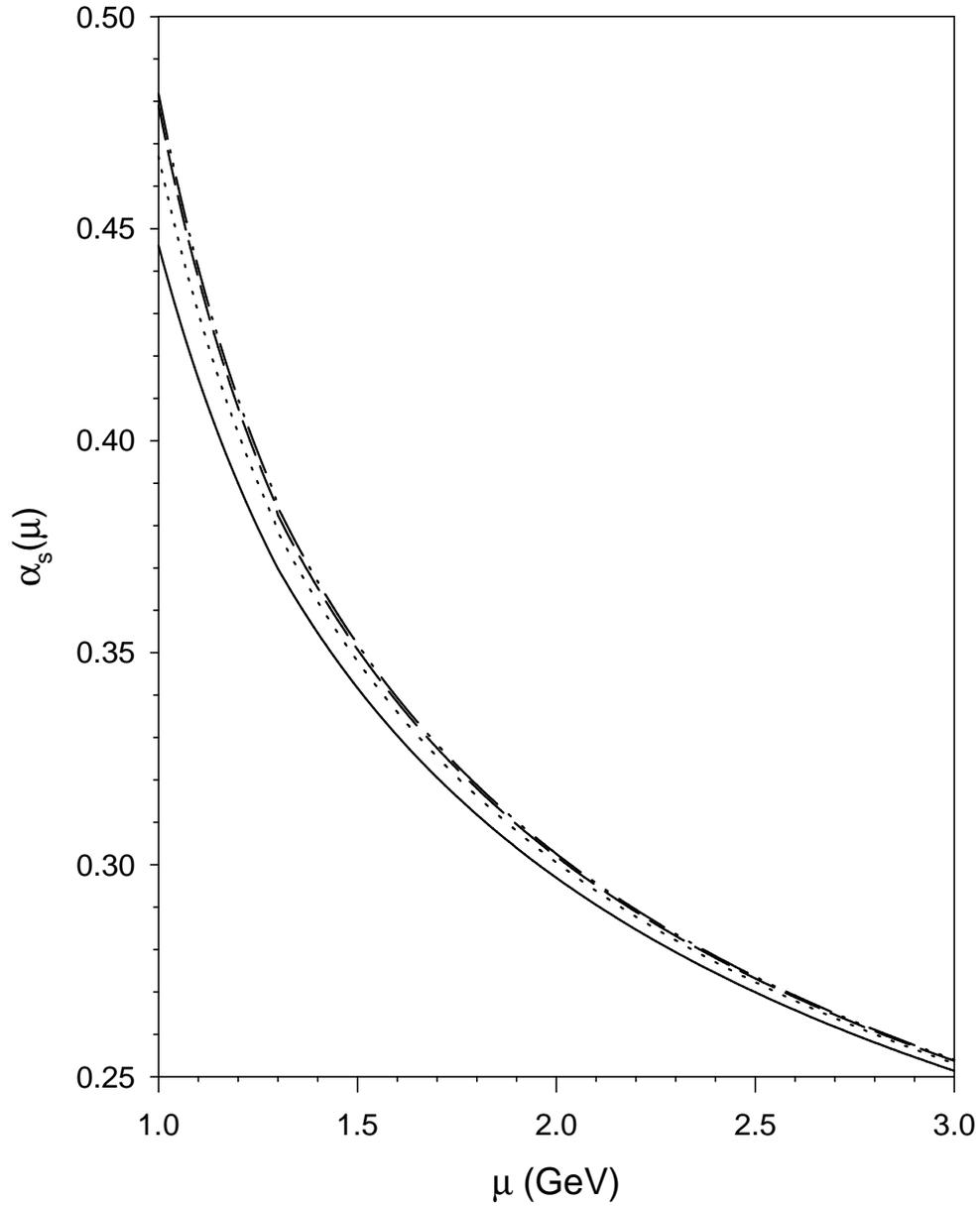}
\caption{
Effect of increasing the order of perturbation theory 
in the QCD evolution of the strong coupling constant, using 
$\alpha_s\left(M_Z\right)$ as an initial condition.
Higher-loop terms in the $\beta$ function progressively increase
$\alpha_s$ from the 2-loop order bottom (solid) curve to the
Pad\'e summation top (dashed-dotted) curve, sandwiching the
three- and four-loop curves. 
}
\label{alphafig}
\end{figure}

The compatibility of the experimentally/phenomenologically determined  values $\alpha_s\left(M_Z\right)$ 
and $\alpha_s\left(M_\tau\right)$ with the QCD evolution of the coupling constant can now be studied.
Figure \ref{alphafig}   shows the effect on $\alpha_s(Q)$ of progressive increases in the number of
perturbative terms in the $\beta$ function, culminating with the Pad\'e summation 
(\ref{beta_pade_nf4},\ref{beta_pade_nf5}) for $\beta$.  It is evident that 
the curves for $\alpha_s(Q)$ appear to converge from below to that generated by the Pad\'e summation
of the $\beta$ function, since the gaps between curves of successive order decrease.
Using the input values (\ref{alpha_Z_pdg},\ref{mb_pdg}) for the QCD evolution of $\alpha_s$ down from the $Z^0$ to $\tau$ mass, 
we obtain the following range of values for $\alpha_s\left(M_\tau\right)$ for successive orders of
perturbation theory \cite{elias}:
\begin{eqnarray}
& &{\rm 2-loop} \qquad 0.3055\le\alpha_s\left(M_\tau\right) \le 0.3383
\label{2loop_alpha_mtau}\\
& &{\rm 3-loop} \qquad 0.3096\le\alpha_s\left(M_\tau\right) \le 0.3442
\label{3loop_alpha_mtau}\\
& &{\rm 4-loop} \qquad 0.3112\le\alpha_s\left(M_\tau\right) \le 0.3468
\label{4loop_alpha_mtau}\\
& &{\rm Pade~ summation} \qquad 0.3119\le\alpha_s\left(M_\tau\right) \le 0.3480
\label{pade_alpha_mtau}
\end{eqnarray} 
The dominant contribution to  the uncertainty (\ref{2loop_alpha_mtau}-\ref{pade_alpha_mtau}) originates from $\alpha_s\left(M_Z\right)$---
the effect of the uncertainty in the five-flavour threshold (\ref{mb_pdg}) is inconsequential.
It is also evident that Pad\'e-improvement via (\ref{beta_pade_nf4}) and (\ref{beta_pade_nf5}) 
does not alter significantly  the range
of $\alpha_s\left(M_\tau\right)$ devolving from the empirical range for $\alpha_s\left(M_Z\right)$, 
as given in (\ref{alpha_Z_pdg}).  Moreover, the ranges
(\ref{4loop_alpha_mtau}) and (\ref{pade_alpha_mtau}) overlap the lower end of the current 
PDG range (\ref{alpha_tau_pdg}) for the extraction of $\alpha_s\left(M_\tau\right)$
from $R_\tau$. We note, however, that only minimal overlap occurs between (\ref{4loop_alpha_mtau},\ref{pade_alpha_mtau}) 
and the previous (1996)
PDG range  $\alpha_s\left(M_\tau\right) = 0.370 \pm 0.033$ obtained from $R_\tau$ \cite{pdg96}.  
This marginal compatibility with the
RG-devolution estimates of $\alpha_s\left(M_\tau\right)$ provided the original motivation for us to examine the effect
of (estimated) higher-order perturbative corrections on the extraction of
$\alpha_s\left(M_\tau\right)$ from $R_\tau$.

This extraction occurs by comparing the measured value of $R_\tau$ with the calculated value of 
$\delta^{(0)}$,
the purely perturbative QCD corrections to the tree diagram for the calculation of $R_\tau$:
\begin{equation}
R_\tau\equiv \frac{\Gamma\left(\tau\rightarrow\nu_\tau+{\rm hadrons}\right)}{\Gamma\left(\tau\rightarrow\nu_\tau+e+\bar\nu_e\right)}
=3.058\left[1+\delta^{(0)}-\left(0.014\pm 0.005\right)\right]\quad .
\label{R_tau}
\end{equation}
The above expression, as quoted from the current PDG \cite{pdg96}, is based on the seminal analysis of
Braaten, Narison, and Pich \cite{braaten}, as well as more recent work by Neubert \cite{neubert}. In particular, the
numerical factor in parentheses represents non-perturbative contributions that are dominated by
dimension-6 terms.  In the $\overline{\rm MS}$  scheme, the purely perturbative QCD contributions to $R_\tau$ are
\begin{equation}
1+\delta^{(0)}=1+a\left(M_\tau\right)+5.2023 \left[a\left(M_\tau\right)\right]^2
+26.366\left[a\left(M_\tau\right)\right]^3 \quad .
\label{delta_0}
\end{equation}
where $a\left(M_\tau\right)\equiv \alpha_s\left(M_\tau\right)/\pi$.
Using the empirically motivated test value $\alpha_s\left(M_\tau\right) = 0.3500$,
corresponding to the central value of the PDG range (\ref{alpha_tau_pdg}), we find that the 
contributions of
successive terms in (\ref{delta_0}) to $\delta^{(0)}$ are respectively
\begin{equation}
\delta^{(0)}=0.1114+0.06457+0.03646
\end{equation}
illustrating the slow convergence of perturbative field theory at the mass scale $\mu = M_\tau$. The ratio
of successive terms within $\delta^{(0)}$ appears to be nearly 0.6, indicative of significant further
contributions from corrections to (\ref{delta_0}) beyond three-loop-order.

     Asymptotic Pad\'e-approximant methods \cite{samuel2,apap} can be utilized to estimate the aggregate
effect of higher order terms in the series (\ref{delta_0}).  Given a field-theoretical perturbative series of the
form
\begin{equation}
S=1+R_1 a+R_2 a^2+R_3 a^3+R_4 a^4
\label{basic_series}
\end{equation}
in which the coefficients $R_k$ are characterized by asymptotic behaviour $R_k\sim k! C^k k^\gamma$
\cite{VZ}, the
coefficient $R_4$, which we assume to be unknown, can be estimated from the known terms $R_1$, $R_2$,
and $R_3$.  An $[N|M]$ Pad\'e-approximant for the series (\ref{basic_series}) then yields coefficients $R_k^{Pade}$ that differ
from those of the series itself via the asymptotic error formula \cite{samuel2,apap}
\begin{equation}
\frac{R^{Pade}_{N+M+1}-R_{N+M+1}}{R_{N+M+1}}\simeq -\frac{M! A^M}{\left[N+M(1+c)+b\right]^M}\quad ,
\label{apap_error}
\end{equation}
where $\{A,c,b\}$ are constants to be determined. For example, a $[2|1]$ Pad\'e approximant to 
(\ref{basic_series})
would lead to the prediction 
\begin{equation}
R^{pade}_4=\frac{R_3^2}{R_2}
\label{2_1_pade}
\end{equation}
It has been shown elsewhere \cite{elias} that the error formula (\ref{apap_error}) can be utilized in conjunction with
$[0|1]$ and $[1|1]$ approximants to determine $A$ and $(c+b)$, thereby leading to the following
``asymptotic Pad\'e-approximant'' (APAP) estimate for $R_4$:
\begin{equation}
R_4^{APAP}=\frac{R_3^2\left[3+(c+b)\right]}{R_2\left[3+(c+b)-A\right]}
=\frac{R_3^2\left[R_2^3+R_1R_2R_3-2R_1^3R_3\right]}{R_2\left[2R_2^3-R_1^3R_3-R_1^2R_2^2\right]}
\label{R4_apap}
\end{equation}
If just the $[2|1]$ Pad\'e-approximant  is used to estimate [via (\ref{2_1_pade})] the $\alpha_s^4$
contribution to $\delta^{(0)}$, it is found that \cite{samuel}
\begin{equation}
1+\delta^{(0)}=1+a\left(M_\tau\right)+5.2023 \left[a\left(M_\tau\right)\right]^2
+26.366\left[a\left(M_\tau\right)\right]^3
+109.2\left[a\left(M_\tau\right)\right]^4
\label{mark}
\end{equation}
an estimate very close to that obtained by Kataev and Starshenko using other methods
\cite{kataev}. The
APAP-estimate of the $\alpha^4$ term, obtained via (\ref{R4_apap}), is also positive and somewhat (20\%) larger: 
\begin{equation}
1+\delta^{(0)}=1+a\left(M_\tau\right)+5.2023 \left[a\left(M_\tau\right)\right]^2
+26.366\left[a\left(M_\tau\right)\right]^3
+132.44\left[a\left(M_\tau\right)\right]^4
\label{trunc}
\end{equation}
It is significant to note that this prediction is very close to the {\em maximum} estimated size 
of the fourth order effect used to determine the theoretical uncertainty in \cite{braaten}, indicating
an underestimate of the higher order effects in the extraction of $\alpha_s\left(M_\tau\right)$ from $R_\tau$.
However, even if $\delta^{(0)}$ includes an estimate of the $\left[a\left(M_\tau\right)\right]^4$ term, one sees from (24) that the
convergence of the perturbative series remains too slow to justify a truncation.  For example,
if $\alpha_s\left(M_\tau\right) = 0.3500$, then the contribution of successive orders to $\delta^{(0)}$ is seen to be
\begin{equation}
\delta^{(0)}=0.1114+0.06457+0.03646+0.02040\quad .
\end{equation}
The (estimated) fourth term is 18\% of the leading perturbative contribution. Such slow
convergence indicates that further higher order terms  should contribute significantly to
$\delta^{(0)}$. A Pad\'e-summation, in this case the $[2|2]$ approximant whose first five Maclaurin expansion
terms replicate the series (\ref{trunc}), provides an estimate of the total effect of higher order terms in
a perturbation series \cite{pade}.  This Pad\'e summation is given by
\begin{equation}
1+\delta^{(0)}=\frac{1-6.5483 a\left(M_\tau\right)+10.5030 \left[a\left(M_\tau\right)\right]^2}{1-7.5483 a\left(M_\tau\right)+12.8514 \left[a\left(M_\tau\right)\right]^2}
\label{pade}
\end{equation}

\begin{figure}
\centering
\includegraphics[scale=0.7]{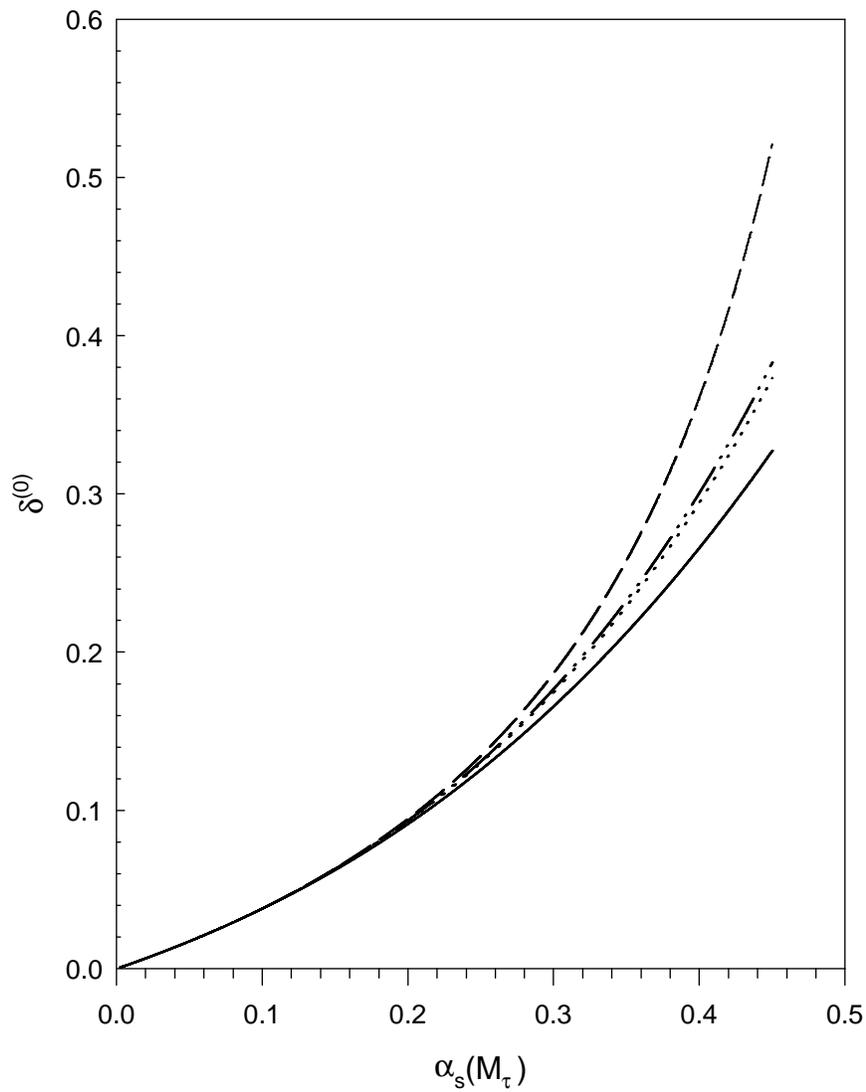}
\caption{ Values for 
$\delta^{(0)}$ as a function of $\alpha_s\left(M_\tau\right)$ 
using 
the four different treatments of $\delta^{(0)}$.  The solid curve uses 
(\protect\ref{delta_0}) [``Truncation],
the dotted curve uses 
(\protect\ref{mark}) [``$[2|1]$''], the dashed-dotted curve uses 
(\protect\ref{trunc}) [``APAP''],
and the dashed curve uses 
(\protect\ref{pade}) [``PS''], as discussed in the text.
}
\label{delta_alpha_fig}
\end{figure}

 Figure \ref{delta_alpha_fig}
compares the
dependence of $\delta^{(0)}$ on $\alpha_s\left(M_\tau\right)$  
obtained from (\ref{delta_0}, \ref{mark}, \ref{trunc}, \ref{pade}).  These curves correspond respectively to
\begin{itemize}
\item {\bf Truncation:} Truncation of contributions to $\delta^{(0)}$ beyond three-loop order (\ref{delta_0});
\item {\bf [2$|$1]:} Inclusion via (\ref{mark}) of the $[2|1]$  Pad\'e-approximant estimate of the four-loop
                  contribution to $\delta^{(0)}$;
\item {\bf APAP:}       Inclusion via (\ref{trunc}) of the asymptotic error-formula estimate of the four-loop
                  contribution to $\delta^{(0)}$;
\item {\bf PS:}         Pad\'e-summation estimate (\ref{pade}) of all higher-loop contributions to $\delta^{(0)}$.
\end{itemize}
Near the PDG value $\alpha\left(M_\tau\right)=0.350$, 
the Pad\'e effects lead to a significant {\em increase} in $\delta^{(0)}$.  In comparison with 
the three-loop perturbative result (\ref{delta_0}), the size of this  enhancement
obtained from the Pad\'e summation (\ref{pade}) is roughly twice the
enhancement obtained by including   Pad\'e estimates of the four-loop  contributions (\ref{trunc},\ref{mark}), indicating the significance of the
higher-order effects estimated in the $[2|2]$ Pad\'e summation.

      In Table \ref{delta_table}, we display the values for  $\alpha_s\left(M_\tau\right)$ one obtains for a given value of $\delta^{(0)}$ by
inverting equations (\ref{delta_0}), (\ref{mark}), (\ref{trunc}) or (\ref{pade}).  
The present empirical range $R_\tau= 3.642 \pm 0.024$ \cite{pdg98}, can be used in conjunction with (\ref{R_tau})
to extract the following range for the purely-perturbative correction $\delta^{(0)}$:
\begin{equation}
\delta^{(0)}=0.2048\pm 0.0129
\label{pdg_delta}
\end{equation}
Using Table \ref{delta_table}, we find 
that this empirical range for $\delta^{(0)}$ determines a corresponding range for 
$\alpha_s\left(M_\tau\right)$ for each case listed above:\footnote{The central value for
``truncation'' increases to $\alpha_s\left(M_\tau\right)=0.347$, consistent with the central value in \protect\cite{pdg98}, provided
we utilize directly the expressions given in \protect\cite{braaten} for the non-perturbative
contributions, which are weakly $\alpha_s$ dependent.}
\begin{eqnarray}
      {\rm Truncation:}\quad \alpha_s\left(M_\tau\right) & = & 0.342 \pm 0.013 
\label{trunc_range}  \\
      \left[2|1\right]: \quad     \alpha_s\left(M_\tau\right) &=& 0.329 \pm 0.011   
\label{21_range}\\
      {\rm APAP:}       \quad \alpha_s\left(M_\tau\right) &=& 0.326 \pm 0.011   
\label{apap_range}\\
      {\rm PS:}      \quad \alpha_s\left(M_\tau\right)    &=& 0.314 \pm 0.010   
\label{ps_range}
\end{eqnarray}
It is evident from the above results that progressively sophisticated Pad\'e-estimates of higher-order corrections
to $\delta^{(0)}$ lead to progressively lower values for $\alpha_s\left(M_\tau\right)$.\footnote{A range similar to (\protect\ref{21_range})
will also follow from the estimate of the $\alpha_s^4$ correction to $\delta^{(0)}$ given in
\protect\cite{kataev}.}
In particular, 
the ranges (\ref{21_range}) and (\ref{apap_range})
are fully enclosed within the ranges (\ref{4loop_alpha_mtau}) and (\ref{pade_alpha_mtau}) from RG-devolution from $\alpha_s\left(M_Z\right)$. 
The range
(\ref{ps_range}), which we regard as the most accurate reflection of cumulative higher order corrections,
is almost entirely enclosed by the lower portion of these RG-ranges. By contrast, only the lower
half of the PDG ``truncation range'' (\ref{alpha_tau_pdg}) for $\alpha_s\left(M_\tau\right)$ is in agreement with the RG-ranges (\ref{4loop_alpha_mtau}) and
(\ref{pade_alpha_mtau}), although the range quoted in (\ref{trunc_range}) does somewhat better than this.  

      We therefore conclude that higher-order corrections to $\delta^{(0)}$ appear to lower the value of
$\alpha_s\left(M_\tau\right)$ extracted from $R_\tau$ by approximately  10\%, but that this lowering seems to improve the
overall compatibility of $\alpha_s\left(M_\tau\right)$ with the QCD evolution of $\alpha_s$ from the present empirical range for $\alpha_s\left(M_Z\right)$. 
Alternatively, one
can conclude that the theoretical uncertainty in $\alpha_s\left(M_\tau\right)$ associated with truncation of contributions
to $\delta^{(0)}$ past three-loop order is not likely to be bi-directional, as indicated in \cite{pdg98}, but is rather an
${\cal O}(10\%)$ effect in the downward direction.

\noindent
{\bf Acknowledgment:} We are grateful for financial support from the Natural Sciences and
Engineering Research Council of Canada.

\clearpage
\begin{table}
\centering
\begin{tabular}{||c|c|c|c|c||}\hline\hline
$\delta^{(0)}$ & \multicolumn{4}{c||}{$\alpha_s\left(M_\tau\right)$ }\\
 & Truncation  & $[2|1]$ & APAP & PS \\\hline\hline
 \hline
           .190& .3268045079& .3148226450& .3126282933& .3025559995
\\ \hline
            .191& .3278720443& .3157796995& .3135669239& .3033691708
\\ \hline
            .192& .3289362783& .3167332605& .3145020527& .3041782424
\\ \hline
            .193& .3299972341& .3176833554& .3154337071& .3049832461
\\ \hline
            .194& .3310549354& .3186300106& .3163619139& .3057842132
\\ \hline
            .195& .3321094053& .3195732518& .3172867002& .3065811741
\\ \hline
            .196& .3331606671& .3205131058& .3182080919& .3073741596
\\ \hline
            .197& .3342087438& .3214495976& .3191261155& .3081632002
\\ \hline
            .198& .3352536583& .3223827527& .3200407965& .3089483254
\\ \hline
            .199& .3362954327& .3233125966& .3209521606& .3097295647
\\ \hline
            .200 & .3373340894 & .3242391534 & .3218602326 & .3105069480
\\ \hline
            .201 & .3383696501 & .3251624483 & .3227650377 & .3112805039
\\ \hline
            .202 & .3394021367 & .3260825052 & .3236666006 & .3120502613
\\ \hline
            .203 & .3404315704 & .3269993482 & .3245649460 & .3128162486
\\ \hline
            .204 & .3414579730 & .3279130010 & .3254600973 & .3135784937
\\ \hline
            .205 & .3424813650 & .3288234873 & .3263520787 & .3143370249
\\ \hline
            .206 & .3435017672 & .3297308298 & .3272409138 & .3150918692
\\ \hline
            .207 & .3445192005 & .3306350518 & .3281266258 & .3158430544
\\ \hline
            .208 & .3455336849 & .3315361762 & .3290092380 & .3165906074
\\ \hline
            .209 & .3465452410 & .3324342253 & .3298887731 & .3173345551
\\ \hline
            .210 & .3475538883 & .3333292210 & .3307652536 & .3180749239
\\ \hline
            .211 & .3485596462 & .3342211861 & .3316387021 & .3188117395
\\ \hline
            .212 & .3495625353 & .3351101415 & .3325091400 & .3195450287
\\ \hline
            .213 & .3505625736 & .3359961093 & .3333765900 & .3202748172
\\ \hline
            .214 & .3515597814 & .3368791108 & .3342410733 & .3210011296
\\ \hline
            .215 & .3525541766 & .3377591664 & .3351026111 & .3217239918
\\ \hline
            .216 & .3535457785 & .3386362978 & .3359612251 & .3224434288
\\ \hline
            .217 & .3545346055 & .3395105254 & .3368169358 & .3231594650
\\ \hline
            .218 & .3555206760 & .3403818694 & .3376697639 & .3238721252
\\ \hline
            .219 & .3565040080 & .3412503501 & .3385197303 & .3245814337
\\ \hline
            .220 & .3574846194 & .3421159877 & .3393668551 & .3252874143
\\\hline\hline
\end{tabular}
\caption{
Values  of $\alpha_s\left(M_\tau\right)$  for given values of
$\delta^{(0)}$ obtained by inverting
(\protect\ref{delta_0}) [``Truncation],
(\protect\ref{mark}) [``$[2|1]$''],  
(\protect\ref{trunc}) [``APAP''],
and 
(\protect\ref{pade}) [``PS''] as discussed in the text.
}
\label{delta_table}
\end{table}

\clearpage

\clearpage

\end{document}